# Polaron transport of amorphous semiconductors with embedded crystallites

David Emin
Department of Physics and Astronomy
University of New Mexico
Albuquerque, New Mexico 87131 USA

Abstract

Near-room-temperature (narrow-band) polaron transport of an amorphous semiconductor with embedded annealing-induced semiconducting crystallites is treated within an effective-medium approach. Carrier mobilities in the crystallites are assumed much larger than those of the amorphous phase. Nonetheless, crystallites act as macroscopic traps when their carriers' energies lie below those in the amorphous phase. Then the mixture's dc conductivity falls below that of the amorphous phase at low enough carrier concentrations. However, with increasing carrier concentration the shifting chemical potential diminishes this trapping effect, enabling crystallites' larger mobilities to drive the mixtures' electrical conductivity above that of the amorphous phase. Meanwhile the Seebeck coefficient remains insensitive to the annealing-induced introduction and growth of embedded crystallites. These features are qualitatively similar to those reported for an amorphous organic polymer FET with annealing-induced embedded crystallites.

## I. INTRODUCTION

Effective-medium approaches have been utilized since the 19$^{th}$-century to address the properties of mixtures. For example, the electrical resistivity of a mixture is then a smooth function of the resistivity of each of its components. More generally, effective-medium methods describe the transport coefficients of a mixture as weighted averages of the transport coefficients of its constituents.

Electronic transport in solids corresponds to flow through a network of circuit elements. For example, the conductivity of electronic carriers executing thermally assisted hops among impurity states in lightly doped semiconductors corresponds to motion through a network of inequivalent resistors [1]. An effective-medium analysis of this circuit consists of averaging of parallel series of resistances [1].

This effective-medium treatment of hopping conduction fails in the extreme low-temperature limit when the relative variations of the resistances associated with hopping becomes very large ($> 10^7$ - $10^8$) [2]. Then percolation phenomena (e.g. variable-range hopping) manifest themselves [3,4]. Unlike effective-medium electronic transport, percolation transport is characterized by abrupt changes with composition (e.g. "percolation thresholds") resulting from tortuous constricted paths that are limited by "critical hops."

The present work uses an effective-medium approach to address electronic transport near room-temperature of a mixture comprising annealing-induced crystallites embedded within an amorphous phase of the same material. As described by "nucleation and growth," crystallites grow from nucleated seeds by progressively incorporating amorphous material on their ever enlarging surface areas. Moreover, as observed, the distribution of sizes of embedded crystallites is usually not that of a random mixture since the energetic stability of bulk crystals relative to that of their amorphous counterparts drives larger crystallites to form at the expense of smaller crystallites (e.g. Ostwald ripening) [5,6].

Over their long history, effective-medium models have been utilized to address the electrical resistivity of very different types of binary mixtures [7-12]. Prior works, like the present paper, consider ordered arrangements of regularly shaped (spheres, cylinders, and cubes) foreign material encased within a host material [7,8]. Since foreign additions are all surrounded by contiguous host material, the mixture's resistivity is not qualitatively modified by the incorporation of embedded material unless its concentration is very large [8,11]. For example, Odelevsky's effective-medium treatments finds that even replacing half of a material's volume with embedded cubes only 1) decreases the resistivity by 75% when the cubes have infinite conductivity and 2) increases the resistivity by 5/2 when the cubes are voids [8,11].

By contrast, other studies consider the resistivity of random mixtures of two components [9,10]. These models generate intertwined contiguous networks of each component. The resulting electrical transport tends to be dominated by the mixture's more conductive component with a resistivity that is a strong non-linear function of its components fractional volumes. For example, Landauer's effective-medium treatment of a 50/50 mixture of conducting and insulating elements finds its conductivity to be one-quarter that of the pure conductor [9].

Here the embedded crystallites are presumed to manifest high-mobility electronic transport. By contrast, the amorphous material is associated with low-mobility electronic transport. For example, near room-temperature, mobilities of large polarons in crystalline semiconductors are generally greater than 1 cm$^2$/V-sec (but usually less than 100 cm$^2$/V-sec) while mobilities of hopping small polarons in amorphous semiconductors are generally less than 1 cm$^2$/V-sec (but often > $10^{-3}$ cm$^2$/V-sec) [13].

This paper begins in Sec. II with a description of cubes of crystallite encased within a matrix of its amorphous counterpart. In Sec. III the net resistance and the Seebeck coefficient of this mixture are described in terms of a network of resistors and EMFs. In the effective-medium regime, the net resistance and Seebeck coefficient of this network reduces to that of a homogeneous medium with suitably averaged transport parameters. In Sec. IV expressions for the resistances and Seebeck coefficients of crystallites and amorphous material whose carriers are (narrow-band) polarons are incorporated into these formulae to yield the resistance and Seebeck coefficient of our binary mixture. Subsections IVA and IVB contain evaluation and analysis of these formulae in two models where crystallites comprise half of the material's volume. In the first example half of the cells are filled with crystallites while the second example takes all cells to be half-filled by crystallites. The paper concludes in Sec. V with a brief summary and a discussion of the relevance of its results to recent transport measurements on an FET comprising an amorphous conjugated polymer with annealing-induced embedded crystallites.

## II. MODEL

This work explicitly addresses transport among a disordered binary arrangement of regions with different mobilities. Figure 1 illustrates a sample of length $NL$ that is divided into cubic cells having sides of length $L$. Most generally only a fraction $f$ of these cubic cells contain cubic crystallites with sides of length $l \equiv gL$ whose carriers move with high mobility. The length of the model's cubes $L$ is chosen so that the crystallites can grow with annealing from their antecedent seeds to fill the cubic cells they occupy as $g$ is increased from 0 to 1. Any annealing-induced increase of the number of crystallites is modelled by increasing $f$. The remaining cubic cells are filled with an amorphous counterpart of the crystalline material. Thus, the fraction of the material's volume occupied by crystallites is just $f_v \equiv fg^3$.

This model corresponds to a network of resistors and EMFs. The evaluation of the transport coefficients of this network is with a generalization of the effective-medium scheme employed to evaluate the resistor-only network of Ref. 1, c.f. its Sec. III.C. Thus, transport is treated as proceeding along parallel paths through differing series of $N$ cells that are generally randomly partially occupied by crystallites. The resistance and Seebeck coefficient for this model are functions of $f$ and $g$.

## III. FORMALISM

The probability of a path passing through $N$ cells of which $n$ are occupied by crystallites and $N - n$ are unoccupied is:

$$P_n(f) \equiv \left[\frac{N!}{n!\,(N-n)!}\right] f^n (1-f)^{N-n}, \quad (1)$$

where $f$ denotes the overall fraction of occupied cells. The binominal theorem ensures normalization of this probability:

$$\sum_{n=0}^{N} P_n(f) = [f + (1-f)]^N = 1. \quad (2)$$

Focusing on *n*, *N* and *N* − *n* being large enough (>> 1) so that their factorials can be evaluated with Stirling's approximation, this distribution function becomes:

$$P_n(f) = \frac{exp\left[-\frac{\left(\frac{n}{N} - f\right)^2}{2f(1-f)/N}\right]}{\sqrt{2\pi n(1 - n/N)}} \rightarrow \frac{exp\left[-\frac{\left(\frac{n}{N} - f\right)^2}{2f(1-f)/N}\right]}{\sqrt{2\pi Nf(1-f)}}. \quad (3)$$

The expression following the arrow indicates that when $P_n(f)$ occurs within integrals over *n*, as it does in Eqs. (6) and (7), its exponential factor restricts *n* to near *fN*. Replacing *n* by *Nf* in the denominator of Eq. (3) results in $P_n(f)$ being simply a normalized Gaussian with *n*/*N* peaked at *f*. Thus, $P_n(f)$ approaches a delta function [$\delta_{n,Nf}$, or $\delta(n - Nf)$ within summations or integrations over *n*, respectively] in the effective-medium limit where the width of this distribution function is ignored.

Current flow cannot avoid the high-resistivity amorphous regions within which embedded crystallites are encased. Thus, except for very large fractional volumes of embedded material, the resistivity of our model mixture does not qualitatively depart from that of the host material (e.g. See Fig. 5 of Ref. 11). In these situations, the resistivity approaches that of a series of resistances from crystallite-containing and purely amorphous cells weighted by their fractional volumes [c.f. See Eq. (1) of Ref. 9 and Fig. 4 of Ref. 11]. As in Sec. III.C of Ref. 1, the net effective-medium resistance of a medium is modelled as a parallel arrangement of such series resistances weighted by $P_n(f)$.

With no boundary resistances, the resistance of an *N*-step *series* containing *n* crystallite-containing cells of resistance $R_c$ and *N* − *n* purely amorphous cells of resistance $R_a$ is

$$R_n = nR_c + (N - n)R_a. \quad (4)$$

Presuming a uniform temperature gradient, the Seebeck coefficient of an *N*-step *series* through *n* crystallite-containing cells with Seebeck coefficient $S_c$ and *N* − *n* cells of amorphous regions with Seebeck coefficient $S_a$ is

$$S_n = [nS_c + (N - n)S_a]/N, \quad (5)$$

since (1) the Seebeck coefficient is just the electro-motive force EMF induced by the application of a temperature differential divided by that temperature difference and (2) the EMF of a series is just the sum of its individual EMFs. Furthermore, the net resistance *R* and Seebeck coefficient of *N*–element *parallel* paths are given by

$$\frac{1}{R} = \sum_{n=0}^{N} \frac{P_n(f)}{R_n} \quad (6)$$

and

$$S = \frac{\sum_{n=0}^{N} \frac{P_n(f)}{R_n} S_n}{\sum_{n=0}^{N} \frac{P_n(f)}{R_n}}. \quad (7)$$

These summations over *n* are evaluated by converting them to integrations with the expressions contained in Eqs. (3-5) incorporated in their integrands. The conductance 1/*R* and Seebeck coefficient *S* are then sums of contributions associated with the Gaussian function's maximum and width, [*f*(1 − *f*)/*N*]:

$$\frac{1}{R} = \frac{1}{N[(1-f)R_a + fR_c]} \left\{ 1 + \left[\frac{(R_c - R_a)}{(1-f)R_a + fR_c}\right]^2 \left[\frac{f(1-f)}{N}\right] \right\} \rightarrow \frac{1}{N[(1-f)R_a + fR_c]} \quad (8)$$

and

$$S = [(1-f)S_a + fS_c]\left\{1 - \left[\frac{(S_c - S_a)}{(1-f)S_a + fS_c}\right]\left[\frac{(R_c - R_a)}{(1-f)R_a + fR_c}\right]\left[\frac{f(1-f)}{N}\right]\right\}$$
$$\rightarrow [(1-f)S_a + fS_c]. \quad (9)$$

As indicated by the expressions following the arrows of Eqs. (8) and (9), simple effective-medium formulae are obtained by ignoring contributions associated with the dispersion parameter, $[f(1 − f)/N]$. Then the transport coefficients, $R$ and $S$, are just averages of those of cells with and without crystallites ($R_c$, $S_c$, $R_a$ and $S_a$) weighted by the volume fractions, $f$ and $1 − f$.

## IV. EFFECTIVE-MEDIUM TRANSPORT COEFICIENTS

Explicit expressions for the effective-medium resistance and Seebeck coefficient, $R$ and $S$ of Eqs. (8) and (9), are now obtained for the binary mixture of crystallite and amorphous semiconductors described in Sec. II and illustrated in Fig. 1. To this end, the resistances of a cell of amorphous semiconducting material $R_a$ and of a cell containing a crystallite $R_c$ are computed first.

The resistance of a cube of amorphous material having sides of length $L$ is $R_a = \rho_a/L$, where $\rho_a \equiv 1/n_a q \mu_a$, the reciprocal of the product of the density of the amorphous material's predominate carriers $n_a$, their charge $q$ and their mobility $\mu_a$. A crystallite within a cell is generally surrounded by amorphous material. The crystallite and the amorphous region that girdles it provide parallel conducting paths with their cross-sectional areas being $l^2$ and $(L^2 − l^2)$, respectively. In addition, the slabs of amorphous material of cross sectional $L^2$ above and below the crystallite and its girdle of amorphous material are connected in series with them. After some algebra, the resistance of a cubic cell having sides of length $L$, whose amorphous regions encapsulate a cubic crystallite having sides of length $l \equiv gL < L$, becomes

$$R_c(g) = \left(\frac{\rho_a}{L}\right)\left[\frac{1 + \left(\frac{\rho_c}{\rho_a} - 1\right)g}{1 + \left(\frac{\rho_c}{\rho_a} - 1\right)g(1 - g^2)}\right], \quad (10)$$

where $\rho_c \equiv 1/n_c q \mu_c$, the reciprocal of the product of the density of a crystallite's predominate carriers $n_c$, their charge $q$ and their mobility $\mu_c$. As the size of the crystallite grows from its antecedent nucleating seed to fill the cell, the cell's resistance $R_c(g)$ changes from $R_c(0) = R_a = \rho_a/L$ to $R_c(1) = \rho_c/L$.

The standard Seebeck coefficient of a cell of an amorphous semiconductor is

$$S_a = \frac{\langle E - \mu \rangle_\sigma}{qT} = \frac{-\mu}{qT}, \quad (11)$$

when the average energy of its carriers is set to zero [14]. The Seebeck coefficient for a cell containing a crystallite is

$$S_c(g) = \frac{-\mu}{qT} + \frac{\Delta}{qT}\left[\frac{g^2}{g^2 + \frac{\rho_c}{\rho_a}(1 - g^2)}\right], \quad (12)$$

where $\Delta$ represents the difference between the average energy of carriers in a crystallite and those in the amorphous phase and cognizance has been taken of the parallel conduction through and around the crystallite. As the size of the crystallite grows from its seed to fill the entire cell, the cell's Seebeck coefficient $S_c(g)$ changes from $S_c(0) = S_a$ to $S_c(1) = (\Delta − \mu)/qT$.

The Seebeck coefficients and the carrier densities depend on the chemical potential $\mu$. An integral relation defines the chemical potential for $n_t$ fermions in terms of the carriers' total density-of-states $\rho(E)$. Materials (e.g. many amorphous, transition-metal-ion and organic semiconductors) whose equilibrated carriers form polarons have very narrow transport bands [13-23]. The chemical potential will then usually reside outside of the transport bands of the amorphous and crystalline components [24]:

$$n_t \equiv \int dE \frac{\rho(E)}{[e^{(E-\mu)/kT} + 1]} \cong \frac{(1 - f_v)\int_{-W_a/2}^{W_a/2} dE \rho_a(E)}{[e^{-\mu/kT} + 1]} + \frac{f_v \int_{\Delta-W_c/2}^{\Delta+W_c/2} dE \rho_c(E)}{[e^{(\Delta-\mu)/kT} + 1]}$$
$$\equiv \frac{(1 - f_v)N_a}{[e^{-\mu/kT} + 1]} + \frac{f_v N_c}{[e^{(\Delta-\mu)/kT} + 1]} = n_a + n_c, \quad (13)$$

where $N_a(T)$ and $N_c(T)$ denote the densities of thermally accessible carrier states of their respective energy bands having widths $W_a$ and $W_c$. These relations become exact in the narrow-band limit, $W_c$ and $W_a \to 0$.

Upon setting $N_c = N_a$ for simplicity and then defining the net carrier concentration $c \equiv n_t/N_c$, the quadratic equation governing the chemical potential becomes:

$$c = \frac{(1-f_v)}{[(e^{-\mu/kT})+1]} + \frac{f_v}{[e^{\Delta/kT}(e^{-\mu/kT})+1]}. \quad (14)$$

Inspection reveals and deeper analysis verifies that the chemical potential varies monotonically with $f_v$:

$$e^{-\mu/kT} \cong \left(\frac{1-c}{c}\right) e^{-f_v \Delta/kT}. \quad (15)$$

The effective-medium conductivity is obtained upon inserting Eq. (10) into Eq. (8):

$$\sigma_{em}(f,g) = \frac{N}{RL} = \frac{1}{L[(1-f)R_a + fR_c]} = \frac{(1-g+g^3) + \left(\frac{\rho_c}{\rho_a}\right)(g-g^3)}{[(1-g+g^3)-fg^3]\rho_a + [(g-g^3)+fg^3]\rho_c}. \quad (16)$$

The effective-medium Seebeck coefficient is obtained from Eq. (9) with use of Eq. (11) and Eq. (12):

$$S_{em}(f,g) = \left(\frac{k}{q}\right)\ln\left(\frac{1-c}{c}\right) + f\left(\frac{\Delta}{qT}\right)\left[\frac{g^2}{g^2 + \frac{\rho_c}{\rho_a}(1-g^2)} - g^3\right]. \quad (17)$$

The components' resistivity ratio is found from their definitions upon using Eq. (15):

$$\frac{\rho_c}{\rho_a} = \left(\frac{\mu_a}{\mu_c}\right) \frac{[(1-c)e^{(1-f_v)\Delta/kT} + c]}{[(1-c)e^{-f_v\Delta/kT} + c]}. \quad (18)$$

These effective-medium formulae are now evaluated for two examples in which crystallites comprise half of the mixture's volume.

## A. HALF OF CELLS FILLED BY CRYSTALLITES: $f = 0.5$, $g^3 = 1$

The expressions for the effective-medium conductivity and Seebeck coefficient, Eqs. (16) and (17), become especially simple in the artificial extreme limiting case where crystallites need not be encased in amorphous material, $g = 1$. Then the ratio of the effective-medium conductivity to that with $n_t$ carriers in the pure amorphous phase is found from Eq. (16) with $g = 1$:

$$\frac{\sigma_{em}(f,1)}{n_t q \mu_a} = \frac{1}{(1-f)[(1-c)e^{-f\Delta/kT}+c] + f(\mu_a/\mu_c)[(1-c)e^{(1-f)\Delta/kT}+c]}$$
$$\rightarrow \frac{1}{(1-f)[(1-c)e^{-f\Delta/kT}+c]}, \quad (19)$$

upon using the definitions of the resistivity of each of the mixture's components and of $n_a$, $n_c$ and $n_t$ contained in and below Eq. (13). The expression after the arrow pertains in the limit of $\mu_a/\mu_c \rightarrow 0$.

Figure 2 shows that for $\Delta > 0$, $\sigma_{em}(0.5,1)/n_t q \mu_a$ of Eq. (19) exceeds unity and falls as $c$ or $T$ is increased. In this case, carriers preferentially occupy the amorphous phase with their electronic transport being constrained by the presence of crystallites. Increasing $c$ or $T$ progressively drives carriers from this conducting channel thereby reducing $\sigma_{em}(0.5,1)/n_c q \mu_a$.

By contrast, as depicted in Fig. 3, with $\Delta < 0$, $\sigma_{em}(0.5,1)/n_t q \mu_a$ of Eq. (19) falls below unity as $c \rightarrow 0$ when the temperature is low enough for $\exp(-f|\Delta|/kT)/(1-f) < 1$. In this situation, carriers preferentially occupy crystallites which function as extended traps that extract carriers from the amorphous region. Increasing $c$ or $T$ promotes carriers from these crystallite traps thereby increasing $\sigma_{em}(0.5,1)/n_t q \mu_a$. Since carriers move through these extended traps with relatively high mobility, $\sigma_{em}(0.5,1)/n_t q \mu_a$ rises above unity when $c$ and $T$ are large enough.

The effective-medium Seebeck coefficient of Eq. (17) with $g = 1$ is unchanged from that of the pure amorphous phase, defined by $g = 0$. That is, $S_{em}(f,g) = (k/q)\ln[(1-c)/c]$ both for $g = 0$ and for $g = 1$. Figure 4 shows the characteristic dependence of $S_{em}$ versus $\ln(c)$. Deviations of the plot of $S_{em}$ versus $\ln(c)$ from linearity occurs at larger values of $c$ than those for which $\sigma_{em}(0.5,1)/n_t q \mu_a$ in Figs. 2 and 3 garner significant dependences on $c$.

## B. ALL CELLS HALF FILLED BY CRYSTALLITES: $f = 1, g^3 = 0.5$

A representative example in which all crystallites are encased in amorphous material is now presented. Here, $g \cong 0.79$, with all cells are half-filled by crystallites. The effective-medium conductivity and Seebeck coefficients of Eqs. (16) and Eq. (17) then become:

$$\frac{\sigma_{em}(1,0.79)}{n_t q \mu_a} = \frac{0.71 + 0.29 \left(\frac{\mu_a}{\mu_c}\right) \frac{[(1-c)e^{\Delta/2kT} + c]}{[(1-c)e^{-\Delta/2kT} + c]}}{0.21[(1-c)e^{-\Delta/2kT} + c] + 0.79 \left(\frac{\mu_a}{\mu_c}\right)[(1-c)e^{\Delta/2kT} + c]}$$

$$\rightarrow \frac{0.71/0.21}{[(1-c)e^{-\Delta/2kT} + c]} \quad (20)$$

and

$$S_{em}(1,0.79) = \left(\frac{k}{q}\right) \ln\left(\frac{1-c}{c}\right) + \left(\frac{\Delta}{qT}\right)\left[\frac{0.63}{0.63 + 0.37\left(\frac{\mu_a}{\mu_c}\right)\frac{[(1-c)e^{\Delta/2kT} + c]}{[(1-c)e^{-\Delta/2kT} + c]}} - 0.5\right]$$

$$\rightarrow \left(\frac{k}{q}\right) \ln\left(\frac{1-c}{c}\right) + \frac{\Delta}{2qT}, \quad (21)$$

where use has been made of Eqs. (15) and (18). The expressions after the arrow result as $\mu_a/\mu_c \rightarrow 0$.

Plots of the relative effective-medium conductivity of Eq. (20) as a function of $c$ and $T$ for $\Delta > 0$ and $\Delta < 0$ shown in Fig. 5 and Fig. 6 are very similar to analogous plots of Eq. (19) displayed in Figs 2 and 3. Furthermore, the plots of Seebeck coefficients versus $c$ shown in Fig. 7 parallel that of Fig. 4. The modest temperature-dependent shifts of the Seebeck coefficients illustrated in Fig. 7 increase in magnitude as the temperature is lowered. However, the effective-medium approach adopted here loses validity as lowering of the temperature increases the disparity between $R_a$ and $R_c$. Furthermore, at low enough temperatures Seebeck coefficients collapse toward zero [24].

## V. SUMMARY AND DISCUSSION

In effective-medium approaches the electronic transport coefficients of a mixture are functions of those of its components. Here the conductivity and Seebeck coefficient of a binary mixture comprising crystallites of a semiconducting material embedded within its semiconducting amorphous phase (depicted in Fig. 1) is addressed.

Restricting the simple effective-medium formulae of Eqs. (8) and (9) to both constituents' electronic carriers exhibiting the narrow-band transport characteristic of polarons yields Eqs. (16) - (18). The carrier mobility of the crystallites is then presumed much greater than that of the amorphous material to generate the plots of Figs. 2-7.

These findings may apply to recent measurements of the conductivity and Seebeck coefficient of the amorphous organic polymer P(NDI20D-T2) before and after the annealing-induced introduction of embedded crystallites [25]. After annealing, crystallites comprise about half of the sample's net volume with the characteristic volume of a crystallite (20 nm × 20 nm × 5 nm = $2 \times 10^{-18}$ cm$^3$) being very much greater than that of one of its monomers, ~$10^{-21}$ cm$^3$.

Transport measurements are performed by incorporating this material in a FET. The conducting channel's carrier density is controlled by altering the FET's gate voltage. The saturation mobility of the amorphous material is < 0.01 cm$^2$/V-s at 180 K and increases as the temperature is raised to 300 K. After annealing the saturation mobility garners a strong dependence on the channel's estimated carrier density. In particular, the ratio of the room-temperature saturation mobility of the annealed material to that of the un-annealed material is about ½ at the lowest measured carrier density ($10^{18}$ cm$^{-3}$) but is about 2 at the highest measured carrier density (~$2 \times 10^{19}$ cm$^{-3}$). In addition, annealing leaves the Seebeck coefficients almost unchanged at a nearly temperature-independent value that is proportional to the logarithm of the carrier density throughout its order-of-magnitude variation.

The concentration dependence of the reported saturation mobility is qualitatively similar to the effective-medium conductivity for $\Delta < 0$ illustrated by the examples plotted in Figs. 3 and 6. Furthermore, the reported Seebeck coefficients resemble those of Figs. 4 and 7. As explained in Sec. IVA, these findings suggest that the crystallites act as extended traps within which carriers move with higher mobility than in the amorphous phase. The increase of the conductivity, proportional to the observed saturation mobility, with increasing carrier concentration results from the carrier concentration $c$ occurring in the denominator of Eq. (19). This dependence on $c$ results from Fermi statistics being employed in Eq. (13). Repulsive interactions are inherent in Fermi statistics since it expresses the Pauli exclusion principle.

The examples illustrated in Figs. 3 and 7 show these repulsive interactions manifesting themselves for $c > 0.1$. Since $c$ is the ratio of the total carrier density $n_t$ to the density of thermally accessible states, the observation (c.f. Fig. 3 of Ref. 11) that the saturation mobility rises above its value in the un-annealed material at $n_t \cong 5 \times 10^{18}$ cm$^{-3}$ implies that the density of thermally accessible states is less than or comparable to $5 \times 10^{19}$ cm$^{-3}$. This value is much less than the total density of transport states, $\sim 10^{21}$ cm$^{-3}$, the density of monomers. In other words, this requirement is that the thermal energy at room temperature, $kT = 0.025$ eV, is much less than the width of the transport polaron band, < the characteristic phonon energy. Since this circumstance is commonly fulfilled, the effective-medium approach developed here appears to offer a plausible qualitative explanation of recent transport data on the mixture comprising annealing-induced crystallites embedded within the amorphous counterpart of P(NDI20D-T2).

In summary, annealing-induced embedded crystallites can act as extended traps when the energies of their carriers lie below those of the surrounding amorphous material. Then the introduction of embedded crystallites lowers carriers' conductivity. As the carrier density is increased, repulsions between crystallites' carriers diminish this trapping effect. Then, with a sufficiently large carrier concentration and the mobility of carriers in crystallites exceeding that of carriers in their amorphous surroundings, the introduction of crystallites increases the conductivity. All told, the introduction of embedded crystallites distinctively changes from lowering the conductivity to increasing the conductivity as the carrier density is increased. This paper only addresses inter-carrier repulsions generated by the Pauli exclusion principle. By contrast, Ref. 25 ignores these repulsions and presumes that its measurements indicate carriers' mutual Coulomb repulsions.

## ACKNOWLEDGEMENTS

The author gratefully acknowledges that this study was motivated by extensive interactions with Henning Sirringhaus and his associates, primarily Martin Statz and Riccardo Di Pietro, who shared their data on the effects of annealing on the field-effect mobility and Seebeck coefficient of films of the partially crystalized polymer P(NDI20D-T2).

FIGURE CAPTIONS

Fig. 1 A matrix composed of cubic boxes of length $L$ (blue) is partially filled by cubic crystallites of length $l$ (red).

Fig. 2 The effective-medium conductivity $\sigma_{em}$ divided by $n_t q \mu_a$, the conductivity of the pure amorphous phase, is plotted versus $c$ for $f = 0.5$, $g = 1$ and $\Delta/kT = 1$, 2 and 3 with $\mu_c/\mu_a$ fixed at 1000.

Fig. 3 The effective-medium conductivity $\sigma_{em}$ divided by $n_t q \mu_a$, the conductivity of the pure amorphous phase, is plotted versus $c$ for $f = 0.5$, $g = 1$ and $\Delta|/kT = -1$, $-2$ and $-3$ with $\mu_c/\mu_a$ fixed at 1000.

Fig. 4 The effective-medium Seebeck coefficient $S_{em}$ in units of $k/q$ is plotted versus $c$.

Fig. 5 The effective-medium conductivity $\sigma_{em}$ divided by $n_t q \mu_a$, the conductivity of the pure amorphous phase, is plotted versus $c$ for $f = 1$, $g = 0.79$ and $\Delta/kT = 1$, 2 and 3 with $\mu_c/\mu_a$ fixed at 1000.

Fig. 6 The effective-medium conductivity $\sigma_{em}$ divided by $n_t q \mu_a$, the conductivity of the pure amorphous phase, is plotted versus $c$ for $f = 1$, $g = 0.79$ and $\Delta/kT = -1$, $-2$ and $-3$ with $\mu_c/\mu_a$ fixed at 1000.

Fig. 7 The effective-medium Seebeck coefficient $S_{em}$ in units of $k/q$ is plotted versus $c$ for $f = 1$, $g = 0.79$ and $\Delta/kT = -1$, $-3$ (red and blue dashed curves) and 1, 3 (red and blue solid curves) with $\mu_c/\mu_a = 1000$.

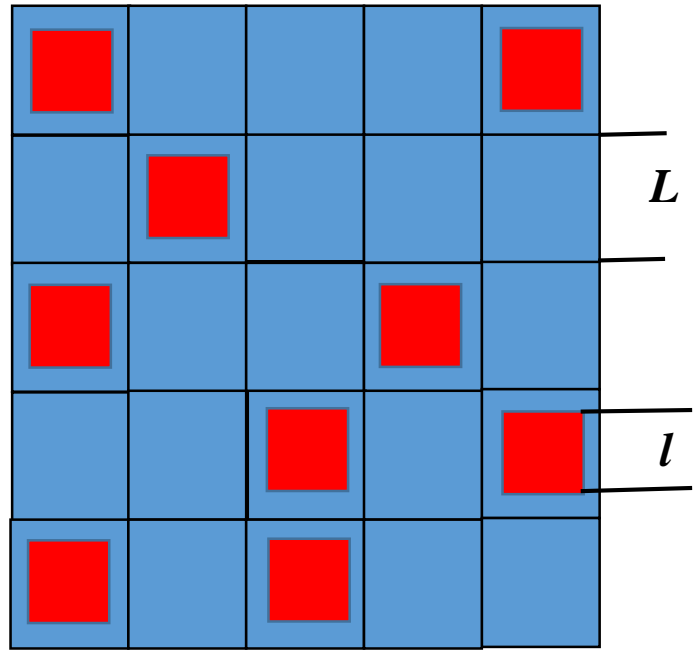

Fig. 1 A matrix composed of cubic boxes of length *L* (blue) is partially filled by cubic crystallites of length *l* (red).

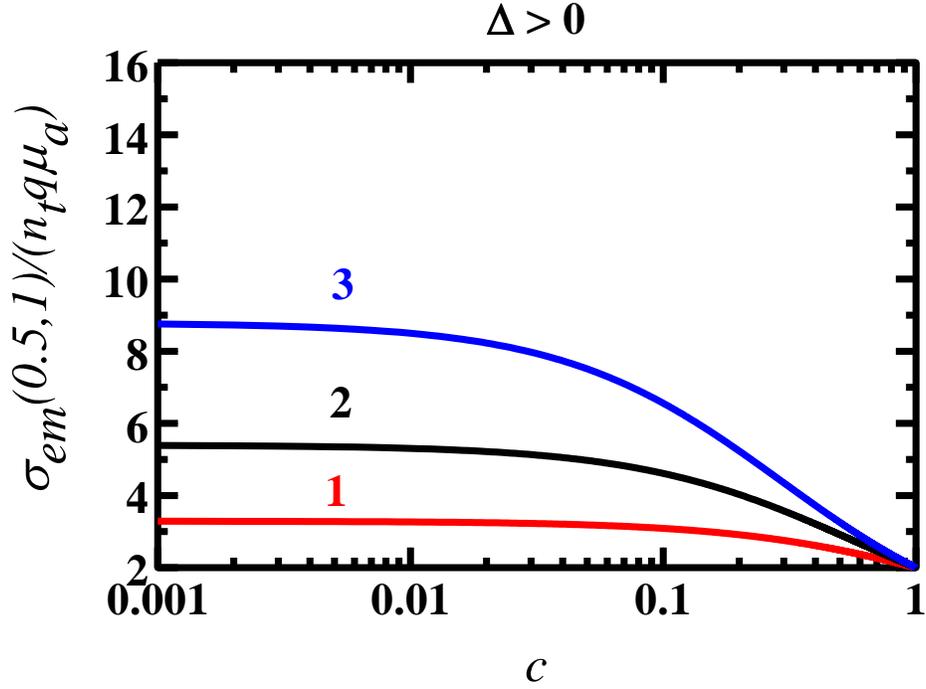

Fig. 2 The effective-medium conductivity $\sigma_{em}$ divided by $n_t q \mu_a$, the conductivity of the pure amorphous phase, is plotted versus $c$ for $f = 0.5$, $g = 1$ and $\Delta/kT = 1$, 2 and 3 with $\mu_c/\mu_a$ fixed at 1000.

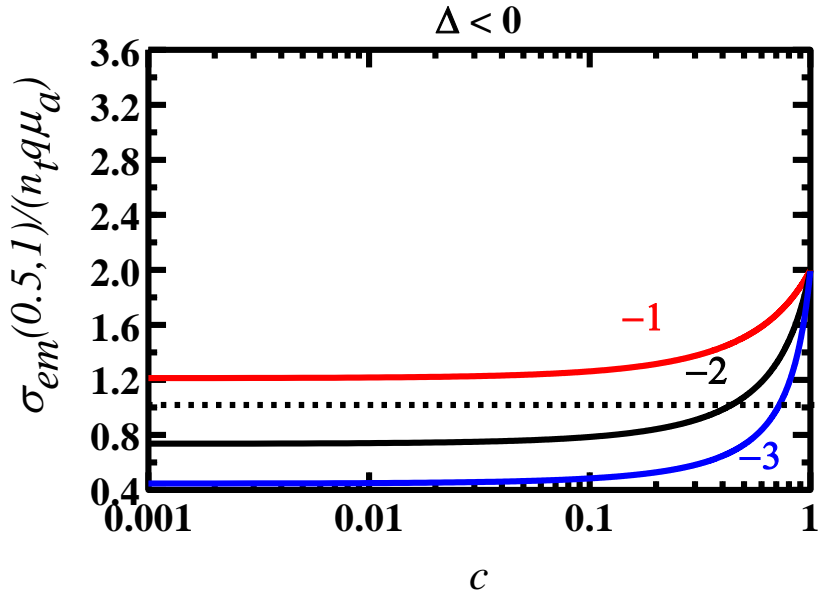

Fig. 3 The effective-medium conductivity $\sigma_{em}$ divided by $n_t q \mu_a$, the conductivity of the pure amorphous phase, is plotted versus $c$ for $f = 0.5$, $g = 1$ and $\Delta/kT = -1$, $-2$ and $-3$ with $\mu_c/\mu_a$ fixed at 1000.

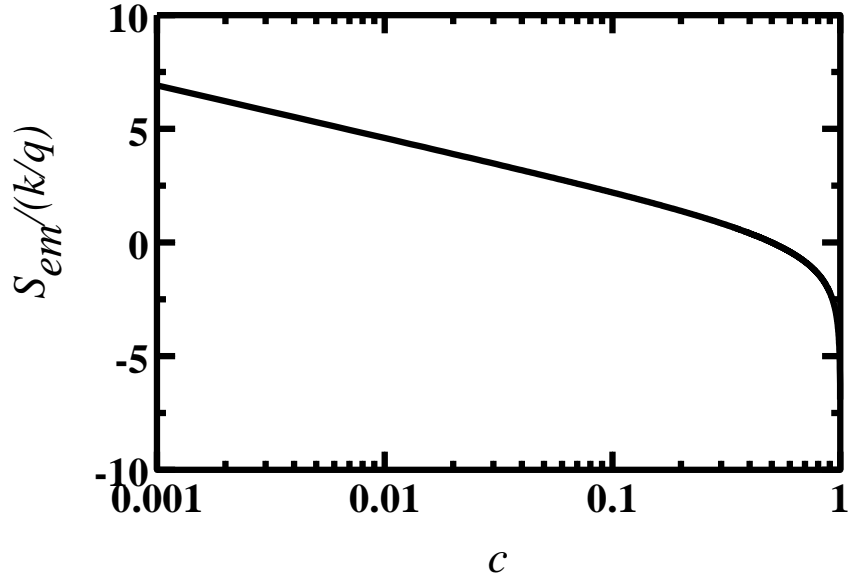

Fig. 4 The effective-medium Seebeck coefficient $S_{em}$ in units of $k/q$ is plotted versus $c$.

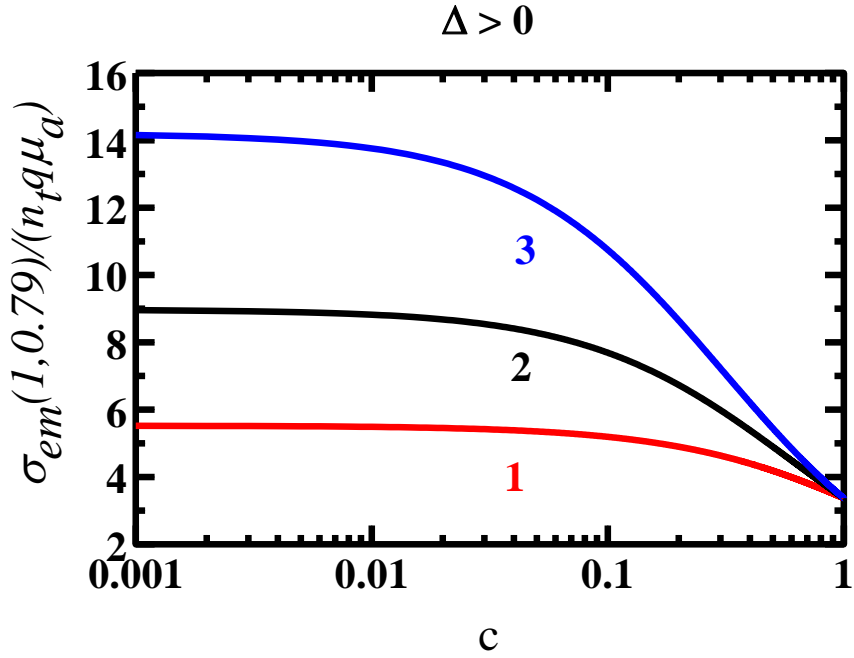

Fig. 5 The effective-medium conductivity $\sigma_{em}$ divided by $n_t q \mu_a$, the conductivity of the pure amorphous phase, is plotted versus $c$ for $f = 1$, $g = 0.79$ and $\Delta/kT = 1$, 2 and 3 with $\mu_c/\mu_a$ fixed at 1000.

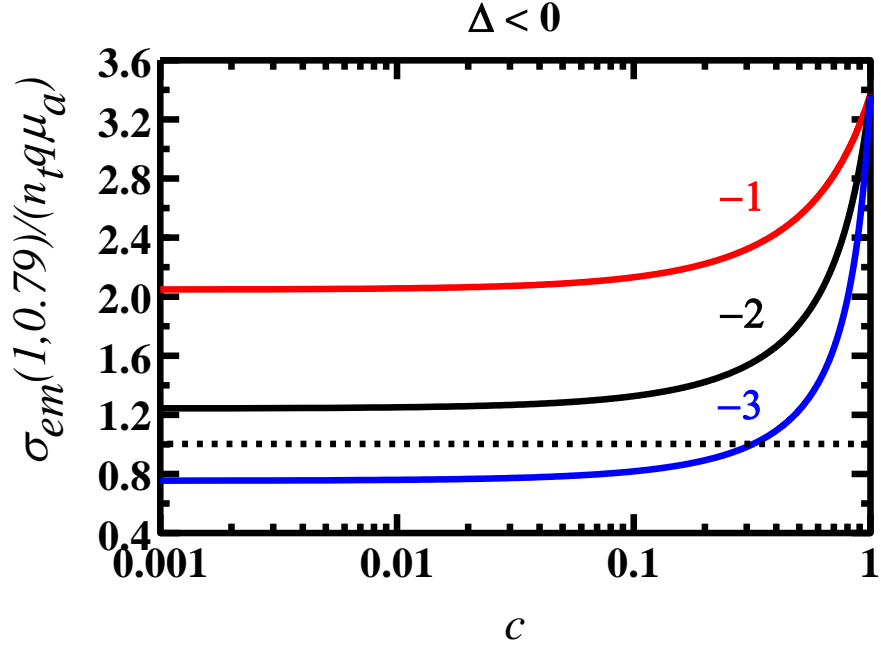

Fig. 6 The effective-medium conductivity $\sigma_{em}$ divided by $n_t q \mu_a$, the conductivity of the pure amorphous phase, is plotted versus $c$ for $f = 1$, $g = 0.79$ and $\Delta/kT = -1, -2$ and $-3$ with $\mu_c/\mu_a$ fixed at 1000.

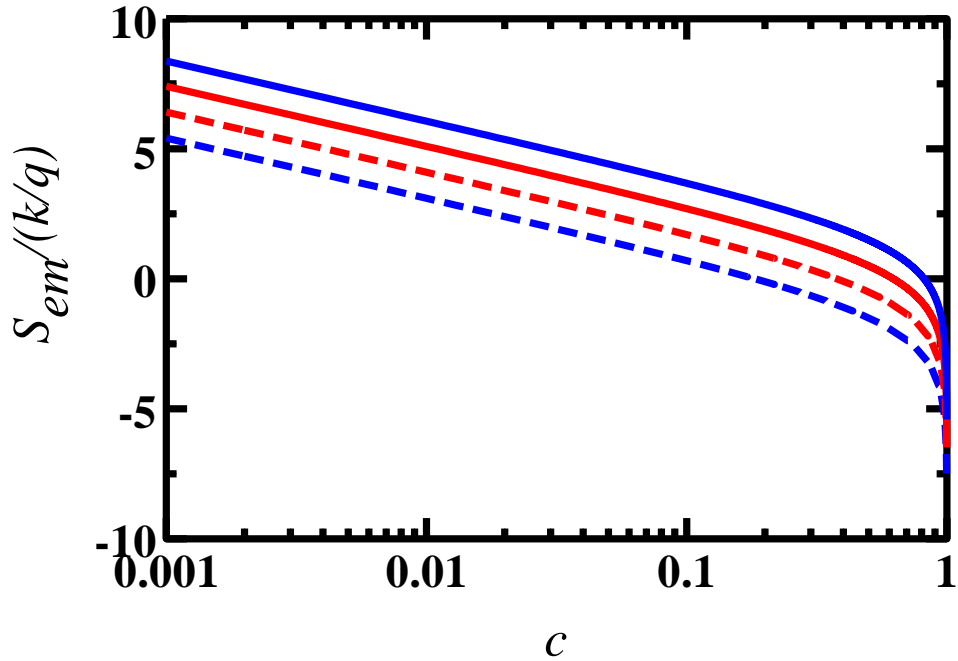

Fig. 7 The effective-medium Seebeck coefficient $S_{em}$ in units of $k/q$ is plotted versus $c$ for $f = 1$, $g = 0.79$ and $\Delta/kT = -1, -3$ (red and blue dashed curves) and $1, 3$ (red and blue solid curves) with $\mu_c/\mu_a = 1000$.